\documentclass[preprint,showpacs,preprintnumbers,amsmath,amssymb]{revtex4}

\usepackage{graphicx}
\usepackage{dcolumn}
\usepackage{bm}

\begin{document}

\title{The role of temperature in the magnetic irreversibility of type-I Pb superconductors}

\author{Sa\"{u}l V\'{e}lez}
\author{Antoni Garc\'{i}a-Santiago}
 \email{agarciasan@ub.edu}
\author{Joan Manel Hernandez}
\author{Javier Tejada}

\affiliation{Grup de Magnetisme, Departament de F\'{i}sica Fonamental, Facultat de F\'{i}sica, Universitat de
Barcelona, c. Mart\'{i} i Franqu\`{e}s 1, planta 4, edifici nou, 08028 Barcelona, Spain
\\
and Institut de Nanoci\`{e}ncia i Nanotecnologia IN$^2$UB, Universitat de Barcelona, c. Mart\'{i} i Franqu\`{e}s 1, edifici nou, 08028 Barcelona, Spain}

\date{\today}

\begin{abstract}
Evidence of how temperature takes part in the magnetic irreversibility in the intermediate state of a cylinder and various disks of pure type-I superconducting lead is presented. Isothermal measurements of first magnetization curves and magnetic hysteresis cycles are analyzed in a reduced representation that defines an equilibrium state for flux penetration in all the samples and reveals that flux expulsion depends on temperature in the disks but not in the cylinder. The magnetic field at which irreversibility sets in along the descending branch of the hysteresis cycle and the remnant magnetization at zero field are found to decrease with temperature in the disks. The contributions to irreversibility of the geometrical barrier and the energy minima associated to stress defects that act as pinning centers on normal-superconductor interfaces are discussed. The differences observed among the disks are ascribed to the diverse nature of the stress defects in each sample. The pinning barriers are suggested to decrease with the magnetic field to account for these results.

\end{abstract}


\maketitle

\section{INTRODUCTION}

The dynamics of normal and superconductor regions in the intermediate state (IS) of type-I superconductors has been studied extensively by means of theoretical and experimental procedures \cite{lan,liv,hue,prov,for1,for2,fra,gol,ceb,men1,gou,men2,pro1,pro2,pro3,pro4,vel,velr,chu}. Magnetic irreversibility has been attributed historically to the presence of either physical or chemical defects in the bulk of the sample \cite{liv}, and to surface and shape effects \cite{prov,for1,for2} that may control both the penetration and the expulsion of the magnetic flux. However, this field has been recently revised on the basis of accurate dc magnetization measurements and high-resolution magneto-optical imaging of the flux patterns in the IS of extremely pure, defect-free type-I lead (Pb) superconductors \cite{pro1,pro2,pro3}. The magnetic irreversibility has been found intrinsic for disks when the applied magnetic field is parallel to the revolution axis, but a fully reversible magnetization has been measured in these samples when the magnetic field is perpendicular to this axis or in ellipsoidal samples in any magnetic field configuration. The irreversibility has been thus ascribed to the presence of a geometrical barrier that depends on the shape of the sample and on the orientation of the applied magnetic field \cite{pro2,vel}. Moreover, it has been observed to be correlated with the formation of different flux patterns depending on which branch is explored along the magnetic hysteresis cycle, being bubbles for flux penetration or labyrinthine structures for flux expulsion \cite{pro2,pro3}. However, the effect of the geometrical barrier vanishes when the applied magnetic field is removed and the system becomes unable to trap any flux line.

This picture is significantly modified when the sample contains other irreversibility sources such as stress defects. In this situation, and only when the magnetic field is parallel to the revolution axis, the defects enhance the capability of the system to trap magnetic flux and, as a result, a large remnant magnetization can be retained at zero field \cite{vel}. The similarity of normal-superconductor interfaces (NSI) in type-I superconductors with domain walls (DW) in ferromagnets led recently to suggest that the effect of the stress defects is to pin the NSI driving the system into a metastable state \cite{chu}. This was tested in a disk of Pb by studying the time-evolution of the remnant magnetization at zero magnetic field. The magnetic relaxation followed a logarithmic dependence which was attributed to the presence of a broad distribution of pinning energy barriers, in good agreement with the nature of the stress defects in this sample. The data obtained at low temperatures were analyzed using a theoretical model that describes the quantum tunneling of interfaces (QTI) mediated by the formation and decay of bumps in the NSI pinned at the defects. Several physical properties of the system, like the average size of the energy barriers, the average characteristic lengths of the bumps, and the crossover temperature between thermal and non-thermal regimes of magnetic relaxation, were thus determined \cite{chu}.

However, the magnetic field dependence of the pinning energy barriers associated with the defects is still far from having been explored. Moreover, to our knowledge, the influence of temperature on either the topological or the pinning contributions to the irreversibility of the magnetic hysteresis cycle has not been reported yet. To investigate these effects, isothermal magnetization curves were measured on different samples at several temperatures and were analyzed in terms of reduced magnitudes defined as $m_r \equiv M/H_c$ and $h \equiv H/H_c$, where $M$ is the measured magnetization, $H$ is the applied magnetic field, $H_c(T) = H_{c0}[1 - (T/T_c)^2]$ is the thermodynamic critical field \cite{tink}, $T$ is the temperature, and $T_c$ is the superconducting transition temperature. In this representation, the resulting $m_r(h,T)$ curves of a sample that presents a reversible magnetization in the whole magnetic field range, like an ellipsoidal sample, should scale onto a single curve, $m_r(h)$, whose theoretical description would only depend on the geometrical factor, $N$ \cite{rose}. The possible temperature dependence of the irreversibility of the system should be therefore reflected on the lack of scaling of the $m_r(h,T)$ curves. A comparative investigation of a cylinder that only exhibits geometrical effects and various disks that also contain structural defects of different nature is made in this work to elucidate the role of temperature in the magnetic irreversibility of these samples.

\section{EXPERIMENTAL SETUP}

We have investigated four samples, labeled $A$ to $D$, all made from a commercial rod of extremely pure lead ($99.999$ at.$\%$) \cite{rod}. Samples $A$ and $B$ are two disks of base area 40 mm$^2$ and thickness 0.2 mm that were made by cold rolling a slice of rod to the desired thickness and cutting out the edges to get an octagonal cross section. Sample $A$ was further annealed for one hour in glycerol at 290$^{\rm{o}}$C in nitrogen atmosphere to reduce the mechanical stress associated to defects. Sample $C$ is also a disk with a shape similar to that of samples $A$ and $B$ but it was heated above the melting point of lead (circa 327.5$^{\rm{o}}$C) and cooled down to room temperature in a few seconds trying to induce stress defects of a nature different than those originated by the cold rolling protocol. Sample $D$ is a cylinder of radius 1.5 mm and length 3 mm that was produced by eroding the rod with a whetstone very slowly to avoid the generation of physical defects in the bulk of the sample.

All magnetic measurements were performed in a commercial superconducting quantum interference device magnetometer with a low-temperature stability better than 0.01 K \cite{qd}. In all cases, $H$ was applied parallel to the axis of revolution ($z$ direction), and isothermal $M(H)$ curves were recorded at constant $T$ values between 2.00 and 6.00 K. The samples were first cooled from 8.00 K (normal state) in zero magnetic field (zfc process) down to the desired $T$ value, and $H$ was then progressively increased up to an intensity slightly above the corresponding $H_c(T)$ value at which $M$ vanished, to obtain the first magnetization curve, $M_{1\rm{st}}(H,T)$. Then $H$ was swept down to $-H_c(T)$ and up to $H_c(T)$ again to obtain respectively the descending and ascending branches, $M_{\rm{des}}(H,T)$ and $M_{\rm{asc}}(H,T)$, of the magnetic hysteresis cycle. The number of experimental points in all the cycles was equal to make sure that the total time needed to record each one was always the same. Fitting the values of $H_c(T)$ obtained from the $M_{1\rm{st}}(H,T)$ curves to the theoretical expression given above produced values of 802 $\pm$ 2 Oe for $H_{c0}$ and 7.23 $\pm$ 0.02 K for $T_c$, in good agreement with standard figures that can be found in the literature for similar samples \cite{landolt}.

\begin{figure}[htbp!]
\includegraphics[scale=0.55]{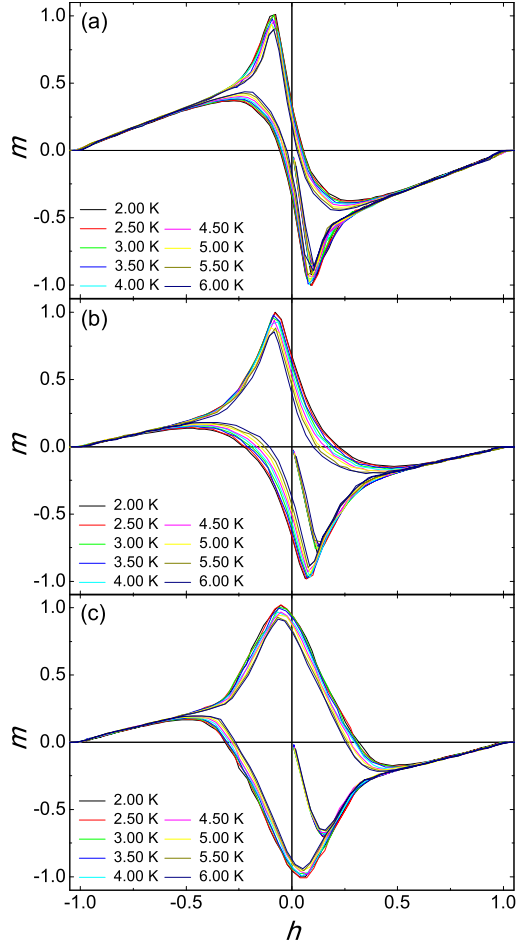}
\caption{(Color online) First magnetization curves and magnetic hysteresis cycles measured on sample $A$ [panel (a)], sample $B$ [panel (b)], and sample $C$ [panel (c)], when the magnetic field is applied parallel to the $z$ direction, after a zfc process down to temperatures that increase progressively between 2.00 K (outermost curve) and 6.00 K (innermost curve) in steps of 0.50 K (see legend for details). The data are plotted using the reduced $m(h)$ representation defined in the text.}
\end{figure}

\section{EXPERIMENTAL RESULTS}

Panels (a), (b), and (c) in Fig. 1 present respectively the magnetization curves for samples $A$, $B$, and $C$ (disks), at temperatures between 2.00 K (outermost curve) and 6.00 K (innermost curve) in steps of 0.50 K, using a representation in which $m_r(h,T)$ has been normalized to the maximum value attained along the cycle at 2.00 K, $m(h,T) \equiv m_r(h,T)/m_{r,\rm{max}}$(2.00 K). The three panels exhibit several common features. First, all the $m_{1\rm{st}}(h,T)$ curves show an initial linear regime (Meissner state) up to a certain value, $h_c' \sim$ 0.1, at which the magnetization enters the intermediate state, goes through a minimum and tends progressively to vanish at $h =$ 1. These curves practically overlap for each sample, namely $m_{1\rm{st}}(h,T) \simeq m_{1\rm{st}}(h)$, suggesting that the process of flux penetration is temperature independent to a large extent. Second, as $h$ decreases from the normal state, a reversible region appears, in which all the $m_{\rm{des}}(h,T)$ branches stick together with $m_{1\rm{st}}(h)$ down to the so-called reduced irreversibility field, $h^*$ \cite{vel}. These branches separate at $h^*$ and stay so until a maximum is reached at a certain $h$ value, below which they tend progressively to merge into a single curve that stretches down to $h = -1$. Third, a substantial amount of trapped flux at zero magnetic field, the so-called reduced remnant magnetization, $m_{\rm{rem}} \equiv m_{\rm{des}}(h = 0)$, is observed. Fourth, as $h$ increases back from $-1$, each $m_{\rm{asc}}(h,T)$ branch turns out to be for the most part antisymmetrical with respect to the corresponding $m_{\rm{des}}(h,T)$ branch, and therefore it reproduces the behavior just described when $h$ evolved in the opposite sense. In particular, each $m_{\rm{asc}}(h,T)$ branch merges again into $m_{\rm{1st}}(h)$ for a $h$ value slightly larger than $h_c'$, that is $m_{\rm{asc}}(h,T) \simeq m_{1\rm{st}}(h)$ from this point on. Finally, both $h^*$ and $m_{\rm{rem}}$ decrease with temperature for each sample and grow from sample $A$ through sample $C$.

Fig. 2 shows the $m(h,T)$ curves obtained for sample $D$ (cylinder) at temperatures between 2.00 and 6.00 K in steps of 1.00 K. There are various details that distinguish this figure from Fig. 1. First, the shape of the hysteresis cycle changes, and in particular the Meissner state extends to larger $h$ values ($h_c' \sim$ 0.6), reflecting the fact that the demagnetizing factor decreases when the thickness of the sample increases ($N \simeq$ 9/10 for the disks \cite{hel}, $N \simeq$ 1/3 for the cylinder \cite{sato}). Second, not only the $m_{\rm{1st}}(h,T)$ curves, but the full hysteresis cycles obtained at different temperatures largely superimpose in the whole magnetic field range, to wit $m(h,T) \simeq m(h)$. Finally, $h^*$ becomes closer to 1 and $m_{\rm{rem}}$ turns almost zero.

\begin{figure}[htbp!]
\includegraphics[scale=0.55]{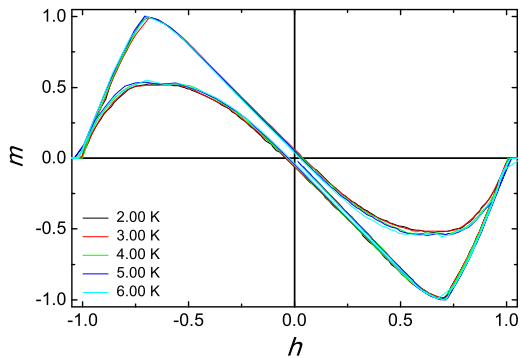}
\caption{(Color online) First magnetization curves and magnetic hysteresis cycles measured on sample $D$, when the magnetic field is applied parallel to the $z$ direction, after a zfc process down to temperatures between 2.00 K and 6.00 K in steps of 1.00 K (see legend for details). The data are plotted using the reduced $m(h)$ representation defined in the text.}
\end{figure}

\section{DISCUSSION}

Let us start discussing the results obtained for sample $D$. First, the fact that $m_{\rm{rem}}$ tends to vanish indicates that the irreversible behavior in this case is only given by the presence of a geometrical barrier \cite{pro2}. This is reinforced by magnetic relaxation experiments performed at several magnetic fields along $m(h)$ that do not show any evolution of the magnetization for times up to two hours, as it is expected for a defect-free sample \cite{vel}. Each magnetic cycle, $M(H,T)$, should then represent the topological hysteresis obtained in this system at the corresponding $T$ value. Moreover, considering that all the cycles scale onto a single $m(h)$ curve in Fig. 2, equivalent spatial distributions of flux patterns (bubbles on penetration, lamellae on expulsion) should form at all temperatures for a given $h$ value, indicating in fact that the topological hysteresis is thermal independent.

Therefore, aside from the variation in the value of $N$ from one sample to another, the differences observed among $A$, $B$ and $C$, and between them and $D$ should be mostly attributed to the presence of structural defects of diverse nature in the disks. In these samples, as the magnetic field is swept, the NSI would move along, they would lose energy due to dissipative processes, and they would be eventually pinned by the defects, leaving the system in a metastable state. Nonetheless, the fact that $m_{\rm{1st}}(h)$ is mostly temperature independent for each sample suggests that the penetration of the magnetic flux from the edges is dominated basically by thermodynamical processes. This means that, even if defects play an important role in the formation of flux patterns \cite{pro3,poo} by shaping superconducting and normal domains via the pinning and depinning of NSI \cite{chu}, their effect at the macroscopic scale would be almost negligible as the magnetic field increases from zero, indicating that $m_{\rm{1st}}(h)$ would represent indeed an equilibrium state for flux penetration into each disk. This is additionally supported by the occurrence of negligible relaxation rates when the time evolution of the magnetization is recorded over two hours after starting at any point along $m_{\rm{1st}}(h)$.

Pinning effects should become thus important in disks during the expulsion of the magnetic flux along $m_{\rm{des}}(h,T)$ and should be therefore invoked to understand both the change in the area of the hysteresis cycles among samples and the thermal dependence of the cycles in each one. The onset of irreversibility at $h^*$ as $h$ decreases along $m_{\rm{des}}(h)$ brings about thermal activation processes that are needed to release the NSI from the pinning by the defects. In analogy with the diffusion of DW in ferromagnets \cite{tej}, the probability that these processes occur is ruled by an Arrhenius equation, $\nu = \nu_0 \exp[-U(h)/k_BT]$, where $\nu_0$ is the so-called attempt frequency, which is usually of order $10^{7}-10^{12}$ Hz, and $U(h)$ is the effective pinning energy barrier, which has been considered to depend actually on $h$. These processes will be relevant at the time scale of our experiments whenever the relation $U(h) \sim 25 k_B T$ is accomplished at a certain $T$ value. Therefore, if this relation is to be satisfied at $h^*$, the decreasing $h^*(T)$ dependence observed in Fig. 1 for all samples would imply that $U(h^*)$ and, in general, $U(h)$ should be a decreasing function as well.

\begin{figure}[htbp!]
\includegraphics[scale=0.55]{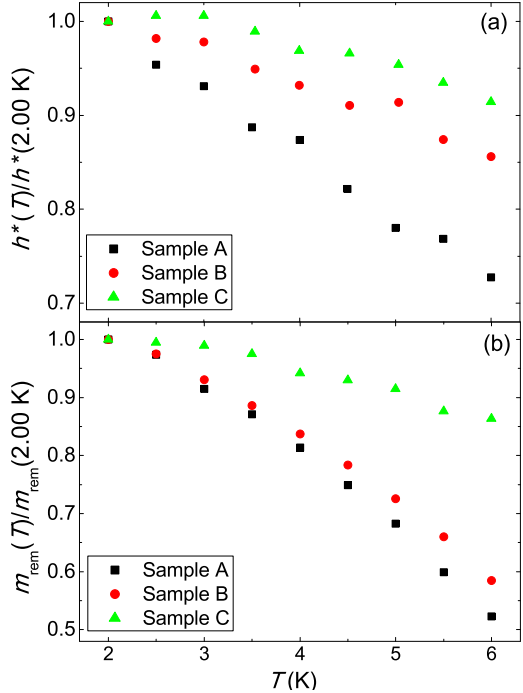}
\caption{(Color online) Temperature dependence of the reduced irreversibility field [panel (a)] and the reduced remnant magnetization [panel (b)], normalized both to their values at 2.00 K, for samples $A$ (squares), $B$ (circles), and $C$ (triangles).}
\end{figure}

The average size of the pinning barriers at zero magnetic field, $U_0 \equiv U(0)$, was recently estimated to be roughly 100 K from magnetic relaxation experiments in a system like sample $A$ \cite{chu}. Similar measurements (not reported here) performed during the current investigation produced values of $U_0$ of approximately 150 and 200 K for samples $B$ and $C$, respectively. Thus, modeling the barrier as $U(h) = U_0 f(h)$, where $f(h)$ contains the functional decreasing dependence of $U(h)$, the range of $h$ at which the relation $U(h) \sim 25 k_B T$ holds at $h^*$ would be simply determined by the actual value of $U_0$. To be precise, as $U_0$ increases progressively from $A$ to $C$, $h^*$ has to increase correspondingly, as it is observed from panel (a) through panel (c) in Fig. 1. Furthermore, this trend is also reflected in the amount of trapped flux at zero magnetic field: as irreversibility sets in at $h^*$, the NSI are immediately pinned by the defects and stay so as $h$ decreases along $m_{\rm{des}}(h)$ down to zero. Therefore, the value of $m_{\rm{rem}}$ will be largely determined by the value of $h^*$, and so it will also increase from $A$ to $C$, as it is shown in Fig. 1. The different size of the pinning energy barriers can be thus held responsible for the change in the area of the whole hysteresis cycle from one sample to another.

We can gain an insight into the thermal dependence of the cycles by looking at how $h^*$ and $m_{\rm{rem}}$ change with temperature. Panels (a) and (b) in Fig. 3 plot respectively $h^*(T)/h^*$(2.00 K) and $m_{\rm{rem}}(T)/m_{\rm{rem}}$(2.00 K) for samples $A$ (squares), $B$ (circles), and $C$ (triangles). All the curves in both panels show a decreasing tendency, with a slope that diminishes from sample $A$ through sample $C$. This is a consequence of the fact that, for a fixed range of $T$ values, the range of $h^*$ values that verify the relation $U(h) \sim 25 k_B T$ has to be reduced as $U_0$ increases from $A$ to $C$, hence the slopes of the $h^*(T)/h^*$(2.00 K) curves decrease in accordance. In addition, as mentioned above, the intensity of $h^*$ establishes the amount of $m_{\rm{rem}}$ that can be trapped at zero field, so the relative variation of $h^*$ with $T$ should also determine the range of values of $m_{\rm{rem}}$ that can be attained at our experimental temperatures. Therefore, $h^*(T)/h^*$(2.00 K) and $m_{\rm{rem}}(T)/m_{\rm{rem}}$(2.00 K) will be correlated for each sample, as it is observed in Fig. 3.

Finally, the reversible region that appears in the cycles at $h>h^*$ could be ascribed in this context to the fact that $U(h)$ might become too small in this range of $h$ to trap magnetic flux by pinning the NSI and, as a consequence, the latter would move quasi freely through the sample \cite{qua}. This would be actually in agreement with previous findings of a decreasing magnetic-field dependence and an upper bound for the velocity at which the superconducting domains move in a current-induced flow experiment in a Pb slab with pinning defects \cite{sol}.

\section{CONCLUSIONS}

To summarize, we have explored the role of temperature in the magnetic irreversibility of a cylinder and various disks of a type-I Pb superconductor when the applied magnetic field is parallel to the axis of revolution of these samples. The magnetic hysteresis cycles were influenced by temperature in the disks, but not in the cylinder. Whereas the contribution of the geometrical barrier was observed to be temperature-independent for all samples, the pinning of the normal-superconductor interfaces in stress defects was found to be responsible for the different thermal dependence of the magnetic hysteresis cycle in each disk. The onset of irreversibility along the descending branch of the cycle was ascribed to the thermal activation of the interfaces over the pinning energy barriers. A decreasing dependence of these barriers on the magnetic field was suggested to explain the fact that both the irreversibility field and the remnant magnetization at zero field diminished as the temperature increases. The variations in the values of these magnitudes and in the whole shape of the hysteresis cycle from one sample to another were attributed to the different size of the pinning energy barriers, which in turn reflects the varied nature of the stress defects created in each sample.

We believe that the results of this paper set the basis for a major exploration of experimental and theoretical questions related to the magnetic irreversibility of type-I superconductors and, in particular, to the role of pinning of normal-superconductor interfaces in the generation of flux patterns. For instance, the hysteresis cycle has been found to be fully reversible in disks when the magnetic field is applied perpendicular to the $z$ direction \cite{vel}, suggesting that pinning effects are not detected when the geometrical barrier does not play a part in the magnetic processes. How this barrier appears to condition the manifestation of pinning during the process of flux expulsion remains an open question yet. On the other hand, the discussion of the magnetic field dependence of the low-temperature relaxation rate of a disk in terms of the QTI model, assuming that the pinning energy barriers decrease with the magnetic field as suggested here, will be the subject of a forthcoming publication \cite{zar}.

\section{ACKNOWLEDGMENTS}

S. V. acknowledges support from Ministerio de Ciencia e Innovaci\'{o}n de Espa\~{n}a. A. G.-S. and J. M. H. thank Universitat de Barcelona for backing their research. J. T. appreciates financial support from ICREA Academia. This work was funded by the Spanish Government project MAT2008-04535.


\begin{thebibliography}{99}
\bibitem{qua} We say "quasi-free motion" at these magnetic field intensities ($h > h^*$) because the magnetic hysteresis cycle is reversible, but the stress defects do still perturbate the NSI and modify the flux patterns in the IS.
\end{thebibliography}
\end{document}